\documentclass[epj]{webofc}
\usepackage[utf8]{inputenc}
\usepackage[varg]{txfonts}
\usepackage{booktabs}
\usepackage{xcolor}
\definecolor{darkred}{rgb}{0.4,0.0,0.0}
\definecolor{darkgreen}{rgb}{0.0,0.4,0.0}
\definecolor{darkblue}{rgb}{0.0,0.0,0.4}
\usepackage[bookmarks,linktocpage,colorlinks,
    linkcolor = darkred,
    urlcolor  = darkblue,
    citecolor = darkgreen]{hyperref}

\usepackage{subfig}
\usepackage{sidecap}

\wocname{EPJ Web of Conferences}
\woctitle{Lattice2017}

\newcommand{à}{\`a}

\newcommand{\be}{\begin{equation}}
\newcommand{\ee}{\end{equation}}
\newcommand{\bea}{\begin{eqnarray}}
\newcommand{\eea}{\end{eqnarray}}

\newcommand{\gdll}{\raisebox{-0.4\totalheight}{\includegraphics[scale=.3]{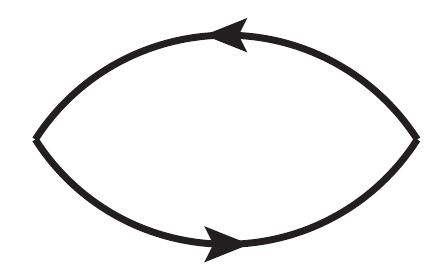}}}
\newcommand{\gdllexch}{\raisebox{-0.4\totalheight}{\includegraphics[scale=.3]{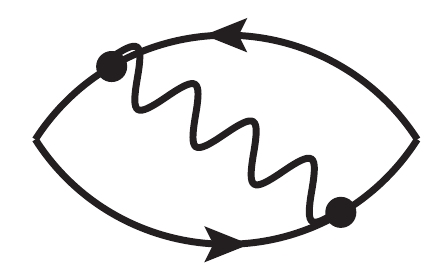}}}
\newcommand{\discgdllexch}{\raisebox{-0.2\totalheight}{\includegraphics[scale=0.4]{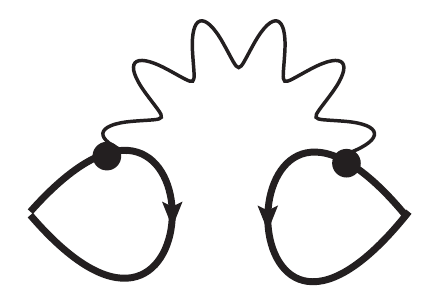}}}
\newcommand{\gdsl}{\raisebox{-0.4\totalheight}{\includegraphics[scale=.3]{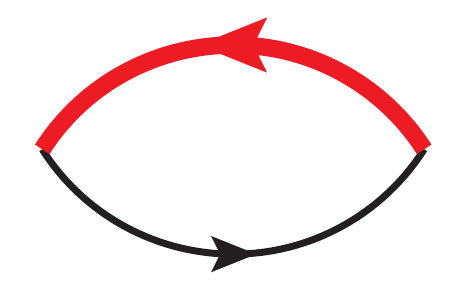}}}
\newcommand{\gdsi}{\raisebox{-0.4\totalheight}{\includegraphics[scale=.3]{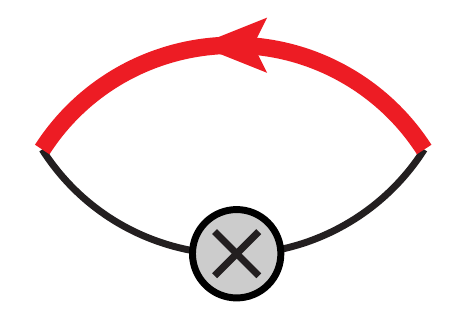}}}

\begin{document}

\selectlanguage{english}

\title{
Leading isospin-breaking corrections to meson masses on the lattice\thanks{{\it Presented at the XXXV International Symposium on Lattice Field Theory, Granada (Spain), 18-24 June 2017.}}
}

\author{
\firstname{Davide} \lastname{Giusti}\inst{1,2} \and
\firstname{Vittorio} \lastname{Lubicz}\inst{1,2} \and
\firstname{Guido} \lastname{Martinelli}\inst{3} \and
\firstname{Francesco} \lastname{Sanfilippo}\inst{2} \and
\firstname{Silvano} \lastname{Simula}\inst{2} \and
\firstname{Nazario} \lastname{Tantalo}\inst{4} \and
\firstname{Cecilia} \lastname{Tarantino}\inst{1,2}
}

\institute{
Dipartimento di Matematica e Fisica, Università degli Studi Roma Tre, Via della Vasca Navale 84, I-00146 Rome, Italy
\and
Istituto Nazionale di Fisica Nucleare, Sezione di Roma Tre, Via della Vasca Navale 84, I-00146 Rome, Italy
\and
Dipartimento di Fisica and INFN, Università degli Studi di Roma "La Sapienza", Piazzale Aldo Moro 5, I-00185 Rome, Italy
\and
Dipartimento di Fisica and INFN, Università degli Studi di Roma "Tor Vergata", Via della Ricerca Scientifica 1, I-00133 Rome, Italy
}

\abstract{
We present a study of the isospin-breaking (IB) corrections to pseudoscalar (PS) meson masses using the gauge configurations produced by the ETM Collaboration with $N_f=2+1+1$ dynamical quarks at three lattice spacings varying from 0.089 to 0.062 fm. Our method is based on a combined expansion of the path integral in powers of the small parameters $(\widehat{m}_d - \widehat{m}_u)/\Lambda_{QCD}$ and $\alpha_{em}$, where $\widehat{m}_f$ is the renormalized quark mass and $\alpha_{em}$ the renormalized fine structure constant. We obtain results for the pion, kaon and $D$-meson mass splitting; for the Dashen's theorem violation parameters $\epsilon_\gamma(\overline{\mathrm{MS}}, 2~\mbox{GeV})$, $\epsilon_{\pi^0}$, $\epsilon_{K^0}(\overline{\mathrm{MS}}, 2~\mbox{GeV})$; for the light quark masses $(\widehat{m}_d - \widehat{m}_u)(\overline{\mathrm{MS}}, 2~\mbox{GeV})$, $(\widehat{m}_u / \widehat{m}_d)(\overline{\mathrm{MS}}, 2~\mbox{GeV})$; for the flavour symmetry breaking parameters $R(\overline{\mathrm{MS}}, 2~\mbox{GeV})$ and $Q(\overline{\mathrm{MS}}, 2~\mbox{GeV})$ and for the strong IB effects on the kaon decay constants.
}

\maketitle

\section{Introduction}
\label{intro}

In the last few years the determination of several observables in flavour physics by lattice QCD reached such a precision that both electromagnetic (e.m.) effects and strong isospin breaking corrections, generated by the light-quark mass difference $(\widehat{m}_d - \widehat{m}_u)$, cannot be neglected any more (see e.g.~Ref.~\cite{FLAG} and references therein).  
Typical examples are the calculations of the leptonic decay constants $f_K$ and $f_\pi$ relevant for $K_{\ell 2}$ and $\pi_{\ell 2}$ decays, and the determination of the vector form factor at zero four-momentum transfer $f_+(0)$ appearing in semileptonic $K_{\ell 3}$ decays. 
These quantities are used to extract the CKM entries $\vert V_{us} \vert$ and $\vert V_{us}\vert / \vert V_{ud} \vert$ from the experimental decay rates, and they have been computed on the lattice with a precision at the few per mille level~\cite{FLAG}.
Such a precision is of the same order of the uncertainties of the e.m.~and strong IB corrections to the leptonic and semileptonic decay rates~\cite{FlaviaNet}. 

The issue of how to include electromagnetic effects in the hadron spectrum and in the determination of quark masses from {\it ab-initio} lattice calculations was addressed for the first time in Ref.~\cite{Duncan:1996xy}. 

Till now the inclusion of QED effects in lattice QCD simulations has been carried out following mainly two methods: in the first one QED is added directly to the action and QED+QCD simulations are performed at few values of the electric charge (see, e.g., Ref.~\cite{Borsanyi:2014jba,Boyle:2016lbc}), while the second one, the RM123 approach of Ref.~\cite{deDivitiis:2011eh,deDivitiis:2013xla}, consists in an expansion of the lattice path-integral in powers of the two {\it small} parameters $(\widehat{m}_d - \widehat{m}_u)$ and $ \alpha_{em}$, namely $\alpha_{em} \approx (\widehat{m}_d - \widehat{m}_u) / \Lambda_{QCD} \approx 1 \%$.
Since it suffices to work at leading order in the perturbative expansion, the attractive feature of the RM123 method is that the small values of the two expansion parameters are factorized out, so that one can get relatively large numerical signals for the {\it slopes} of the corrections with respect to the two expansion parameters. 
Moreover the slopes can be determined using isospin symmetric QCD gauge configurations.

Using the gauge ensembles generated by the European Twisted Mass Collaboration (ETMC) with $N_f = 2 + 1 + 1$ dynamical quarks \cite{Baron:2010bv,Baron:2011sf}, we have calculated the pion, kaon, charmed-meson mass splittings and various $\epsilon$ parameters describing the violations of the Dashen's theorem \cite{Dashen:1969eg} (see Ref.~\cite{FLAG}) by adopting the RM123 method within the quenched QED approximation.

\section{Simulation details}
\label{sec:simulations}

The gauge ensembles used in this contribution are the ones generated by ETMC with $N_f = 2 + 1 + 1$ dynamical quarks, which include in the sea, besides two light mass-degenerate quarks, also the strange and charm quarks with masses close to their physical values \cite{Baron:2010bv,Baron:2011sf}. 

The lattice actions for sea and valence quarks are the same used in Ref.~\cite{Carrasco:2014cwa} to determine the up, down, strange and charm quark masses in isospin symmetric QCD.
They are the Iwasaki action for gluons and the Wilson Twisted Mass Action for sea quarks.
In the valence sector, in order to avoid the mixing of strange and charm quarks a non-unitary set up was adopted, in which the valence strange and charm quarks are regularized as Osterwalder-Seiler fermions, while the valence up and down quarks have the same action of the sea.
Working at maximal twist such a setup guarantees an automatic ${\cal{O}}(a)$-improvement.

We considered three values of the inverse bare lattice coupling $\beta$ and different lattice volumes.
At each lattice spacing, different values of the light sea quark masses have been considered. 
The light valence and sea quark masses are always taken to be degenerate. 
The bare mass of the strange (charm) valence quark $a \mu_s$ ($a \mu_c$) is obtained, at each $\beta$, using the physical strange (charm) mass and the mass renormalization constants determined in Ref.~\cite{Carrasco:2014cwa}. The values of the lattice spacing are: $a = 0.0885(36), 0.0815(30), 0.0619(18)$ fm at $\beta = 1.90, 1.95$ and $2.10$, respectively.

For each gauge ensemble the pseudoscalar meson masses are extracted from a single exponential fit (including the proper backward signal) in the range $t_{min} \leq t \leq t_{max}$, given explicitly in Ref.~\cite{Giusti:2017dmp}.

Following Refs.~\cite{deDivitiis:2013xla,Gasser:2003hk} we impose a specific matching condition between the full QCD+QED and the isospin symmetric QCD theories: in the $\overline{\mathrm{MS}}$ scheme at a renormalization scale $\mu = 2$ GeV we require $\widehat{m}_f(\overline{\mathrm{MS}}, 2~\mbox{GeV}) = m_f(\overline{\mathrm{MS}}, 2~\mbox{GeV})$ for $f = (ud), s, c$, where $\widehat{m}$  and $m$ are the renormalized quark masses in the full theory and in isosymmetric QCD. A similar condition is imposed on the strong coupling constants of the two theories (i.e.~the lattice spacing). 
These conditions fix the isosymmetric QCD bare parameters and a unique prescription to define the isosymmetric QCD contribution to each hadronic quantity.

\section{Evaluation of the IB corrections}
\label{sec:master}

According to the approach of Ref.~\cite{deDivitiis:2013xla} the e.m.~and strong IB corrections to the mass of a PS meson can be written as
 \be
      M_{PS} = M^{(0)}_{PS} + \left[ \delta M_{PS} \right]^{QED} + \left[ \delta M_{PS} \right]^{QCD} 
      \label{eq:MPS}
 \ee
 with
  \be
        \label{eq:MPS_QED}
        \left[ \delta M_{PS} \right]^{QED} \equiv  4 \pi \alpha_{em} \left[ \delta M_{PS} \right]^{em} + ... ~ ,
  \ee
  \be
        \label{eq:MPS_QCD}
        \left[ \delta M_{PS} \right]^{QCD} \equiv (\widehat{m}_d - \widehat{m}_u) \left[ \delta M_{PS} \right]^{IB} + ... ~ ,
  \ee
where the ellipses stand for higher order terms in $\alpha_{em}$ and $(\widehat{m}_d - \widehat{m}_u)$, while $M^{(0)}_{PS}$ stands for the PS meson mass corresponding to the renormalized quark masses in the isosymmetric QCD theory.
The separation in Eq.~(\ref{eq:MPS}) between the QED and QCD contributions, $\left[ \delta M_{PS} \right]^{QED}$ and $\left[ \delta M_{PS} \right]^{QCD}$, is renormalization scheme and scale dependent \cite{Gasser:2003hk,Bijnens:1993ae}. 

\begin{SCfigure}[1.5][htb!]
\centering
\subfloat[]{\includegraphics[scale=1.0]{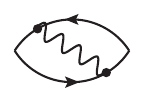}}~
\subfloat[]{\includegraphics[scale=1.0]{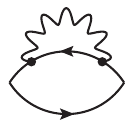}}~
\subfloat[]{\includegraphics[scale=1.0]{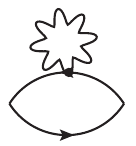}}~
\subfloat[]{\includegraphics[scale=1.0]{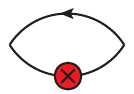}}~
\subfloat[]{\includegraphics[scale=1.0]{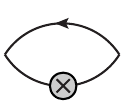}}
\caption{\it \footnotesize Fermionic connected diagrams contributing at $\mathcal{O}(\alpha_{em})$ and $\mathcal{O}(\widehat{m}_d - \widehat{m}_u)$ to the IB corrections to meson masses: exchange (a), self energy (b), tadpole (c), pseudoscalar insertion (d) and scalar insertion (e).}
\label{fig:diagrams}
\end{SCfigure}

Throughout this work we adopt the quenched QED approximation, which neglects the sea-quark electric charges and corresponds to consider only (fermionic) connected diagrams.  
Including the contributions coming from the insertions of the e.m.~current and tadpole operators, of the PS and scalar densities (see Refs.~\cite{deDivitiis:2011eh,deDivitiis:2013xla}) the basic diagrams are those depicted schematically in Fig.~\ref{fig:diagrams}.
The insertion of the PS density is related to the the e.m.~shift of the critical mass present in lattice formulations breaking chiral symmetry, as in the case of Wilson and twisted-mass fermions.

Within the quenched QED approximation the correlator $\delta C^J_{PS}(t)$ corresponds to the sum of the diagrams (\ref{fig:diagrams}a)-(\ref{fig:diagrams}b), while the correlators $\delta C^T_{PS}(t)$, $\delta C^{P_f}_{PS}(t)$ and $\delta C^{S_f}_{PS}(t)$ (where $f = \{u, d, s, c\}$) represent the contributions of the diagrams (\ref{fig:diagrams}c), (\ref{fig:diagrams}d) and (\ref{fig:diagrams}e), respectively.
The removal of the photon zero-mode is done according to $QED_L$~\cite{Hayakawa:2008an}, i.e.~the photon field $A_\mu$ in momentum space satisfies $A_\mu(k_0, \vec{k} = \vec{0}) \equiv 0$ for all $k_0$.

By defining the tree-level correlator $C^{(0)}_{PS}(t)$ as
 \be
     C^{(0)}_{PS}(t) \equiv \sum_{\vec{x}} \langle 0| T \left \{ \phi_{PS}^\dagger(\vec{x}, t) \phi_{PS}(0) \right \} | 0 \rangle ~ ,
     \label{eq:tree}
 \ee
with $\phi_{PS}(x) = i \overline{\psi}_{f_1}(x) \gamma_5 \psi_{f_2}(x)$ being the interpolating field for a PS meson composed by two valence quarks $f_1$ and $f_2$ with charges $q_1 e$ and $q_2 e$, in our analysis the correlators $\delta C^j_{PS}(t)$ with $j = \{J, T, P_f, S_f\}$ are divided by the tree-level one, obtaining at large time distances, where the PS ground-state is dominant,
 \be
      \frac{\delta C^j_{PS}(t)}{C^{(0)}_{PS}(t)} ~ _{\overrightarrow{t >> a, (T-t) >> a}} ~ \frac{\delta Z_{PS}^j}{Z^{(0)}_{PS} } + 
                                                                                                           \frac{\delta M_{PS}^j}{M^{(0)}_{PS}} f_{PS}(t)
      \label{eq:ratio}
 \ee
where $Z^{(0)}_{PS} \equiv \langle 0 | \phi_{PS}(0) | PS^{(0)} \rangle$ and
 \be
      f_{PS}(t) \equiv  M^{(0)}_{PS} \left( \frac{T}{2} - t \right) \frac{e^{- M^{(0)}_{PS} t} - e^{- M^{(0)}_{PS} (T-t)}}{e^{- M^{(0)}_{PS} t} + e^{- M^{(0)}_{PS} (T-t)}} - 
                               1 - M^{(0)}_{PS} \frac{T}{2}
      \label{eq:fPS}
 \ee
is almost a linear function of the Euclidean time $t$.
Thus, the various e.m.~and strong IB corrections to the PS mass, $\delta M_{PS}^j$ ($j = J, T, P_f, S_f$), can be extracted from the slope of the corresponding ratios $\delta C^j_{PS}(t) / C^{(0)}_{PS}(t)$ at large time distances.

\section{Results}
\label{sec:results}

According to Ref.~\cite{deDivitiis:2013xla} the charged/neutral pion mass splitting $M_{\pi^+} - M_{\pi^0}$ is given by
 \be
    M_{\pi^+} - M_{\pi^0} = 4 \pi \alpha_{em} ~ \frac{(q_u - q_d)^2}{2} ~ \partial_t ~ \frac{\gdllexch - \discgdllexch}{\gdll} ~ ,
    \label{eq:pion_splitting}
 \ee
where, following the notation of Ref.~\cite{deDivitiis:2013xla}, ($-\partial_t$) stands for the operator corresponding to the extraction of the slope $\delta M_{PS}$ from the ratio $\delta C_{PS}(t) / C^{(0)}_{PS}(t)$ (see Eq.~(\ref{eq:ratio})).

At first order in the perturbative expansion the pion mass splitting $M_{\pi^+} - M_{\pi^0}$ is a pure e.m.~effect.
Furthermore all the disconnected diagrams generated by the sea quark charges cancel out in the difference $M_{\pi^+} - M_{\pi^0}$ and therefore Eq.~(\ref{eq:pion_splitting}) holds as well in {\it unquenched} QED.
The only remaining {\it disconnected} diagram in Eq.~(\ref{eq:pion_splitting}) is generated by valence quarks in the neutral pion. 
It vanishes in the $SU(2)$ chiral limit~\cite{deDivitiis:2013xla} and, consequently, it is of order of $O(\alpha_{em} m_{\ell})$.
Thus, at the physical pion mass the disconnected contribution to the pion mass splitting $M_{\pi^+} - M_{\pi^0}$ is expected to be a small correction and has been neglected so far in the present study.

Inspired by the Chiral Perturbation Theory (ChPT) analysis of Ref.~\cite{Hayakawa:2008an}, we perform combined extrapolations to the physical pion mass and to the continuum and infinite volume limits, obtaining
\be
    M_{\pi^+} - M_{\pi^0} = 4.21 ~ (23)_{stat+fit} ~ (13)_{syst} ~ \mbox{MeV} = 4.21 ~ (26) ~ \mbox{MeV} ~ ,
    \label{eq:pion_result}
\ee
where $()_{stat+fit}$ indicates the statistical uncertainty including also the ones induced by the fitting procedure and by the errors of the input parameters computed in Ref.~\cite{Carrasco:2014cwa}, namely the values of the average $u/d$ quark mass $m_{ud}$, the lattice spacing and the quark mass RC $1 / Z_P$, while $()_{syst}$ indicates the total systematic uncertainty due to discretization effects, chiral extrapolation and finite volume effects (FVEs).
The determination given in Eq.~(\ref{eq:pion_result}) agrees with the experimental determination
\be
    \left[ M_{\pi^+} - M_{\pi^0} \right]^{exp} = 4.5936 ~ (5) ~ \mbox{MeV}
    \label{eq:pion_exp}
\ee
within $\approx 1.5$ standard deviations.
The difference among the central values, which is equal to $\approx 8 \%$, may be of statistical origin, but it may be due also to the disconnected contribution at order ${\cal{O}}(\alpha_{em} m_{\ell})$ in Eq.~(\ref{eq:pion_splitting}) as well as to possible higher-order effects proportional to $\alpha_{em} (\widehat{m}_d - \widehat{m}_u)$ and to $(\widehat{m}_d - \widehat{m}_u)^2$, which have been neglected.
The latter ones are estimated to be of the order of $\simeq 4 \%$ in Ref.~\cite{FLAG} and therefore the disconnected contribution at order ${\cal{O}}(\alpha_{em} m_{\ell})$ is expected to be of the same size $\approx 4 \%$, which corresponds to $\approx 0.2$ MeV.
 
The Dashen's theorem~\cite{Dashen:1969eg} states that in the chiral limit the self-energies of the neutral Nambu-Goldstone bosons vanish. 
Thus, the violation of the Dashen's theorem in the pion sector can be measured through the quantity $\epsilon_{\pi^0}$ defined as~\cite{FLAG}
\be
    \epsilon_{\pi^0}  = \left[\delta M^2_{\pi^0} \right]^{QED} {\Large \mbox{/}} ~ \left( M_{\pi^+}^2 - M_{\pi^0}^2 \right) ~ .
    \label{eq:epsilon_pi0}
\ee
At the physical pion mass and in the continuum limit we obtain
\be
     \epsilon_{\pi^0} = 0.028 ~ (3)_{stat+fit} ~ (4)_{syst} ~ (44)_{qQED} = 0.028 ~ (44) ~ ,
     \label{eq:epsilon_pi0_result}
 \ee
which is consistent with the FLAG estimate $\epsilon_{\pi^0} = 0.07 ~ (7)$~\cite{FLAG}, based on the old determination of Ref.~\cite{Duncan:1996xy} (corrected by FLAG into the value $\epsilon_{\pi^0} = 0.10 ~ (7)$) and on the more recent result $\epsilon_{\pi^0} = 0.03 ~ (2)$ obtained by the QCDSF/UKQCD collaboration~\cite{Horsley:2015vla}.
In Eq.~(\ref{eq:epsilon_pi0_result}), the $()_{stat+fit}$ and $()_{syst}$ error budgets are estimated as in the case of the pion mass splitting, while $()_{qQED}$ represents our estimate of the uncertainty related to the neglect of the neutral pion disconnected diagram.

The Dashen's theorem predicts that in the chiral limit the e.m.~corrections to the charged kaon and pion are equal to each other, while the ones for the neutral mesons are vanishing.
Therefore, the violation of the Dashen's theorem is parameterized in terms of the quantity $\epsilon_\gamma$ defined as~\cite{FLAG}
\be
     \epsilon_\gamma (\overline{\mathrm{MS}}, \mu) = \frac{\left[ M_{K^+}^2 - M_{K^0}^2 \right]^{QED}(\overline{\mathrm{MS}}, \mu)}{M_{\pi^+}^2 - M_{\pi^0}^2} - 1 ~ .
     \label{eq:epsilon_gamma}
\ee

Within the quenched QED approximation, at the physical pion mass and in the continuum and infinite volume limits our result for the QED contribution to the kaon mass splitting in the $\overline{\mathrm{MS}}$ scheme at a renormalization scale equal to $\mu = 2$ GeV is
\be
    \left[ M_{K^+} - M_{K^0} \right]^{QED} (\overline{\mathrm{MS}}, 2~\mbox{GeV}) = 2.07 ~ (10)_{stat+fit} ~ (5)_{syst} ~ (10)_{qQED} ~\mbox{MeV} = 2.07 ~ (15) ~\mbox{MeV} ~ ,
    \label{eq:kaon_qed_result}
\ee
where $()_{qQED}$ is the estimate of the effects due to the quenched QED approximation ($5 \%)$ taken from Refs.~\cite{Portelli:2012pn,Fodor:2016bgu}.

Using Eqs.~(\ref{eq:pion_result}) and (\ref{eq:kaon_qed_result}) our estimate for $\epsilon_\gamma$ is
 \be
    \epsilon_\gamma(\overline{\mathrm{MS}}, 2~\mbox{GeV}) = 0.801 ~ (48)_{stat+fit} ~ (25)_{syst} ~ (96)_{qQED} = 0.801 ~ (110) ~ ,
    \label{eq:epsilon_g_result}
 \ee
where now the $()_{qQED}$ error includes also the $4 \%$ effect (added in quadrature) coming from the neglect of the neutral pion disconnected diagram. 
Our result (\ref{eq:epsilon_g_result}) is consistent with the FLAG estimate $\epsilon_\gamma = 0.7 ~ (3)$~\cite{FLAG} and with the recent result, converted in the ($\overline{\mathrm{MS}}, 2~\mbox{GeV}$) scheme, $\epsilon_\gamma(\overline{\mathrm{MS}}, 2~\mbox{GeV}) = 0.74 ~ (18)$ from the BMW collaboration~\cite{Fodor:2016bgu} at $N_f = 2+1$ and larger than the recent QCDSF/UKQCD result $\epsilon_\gamma(\overline{\mathrm{MS}}, 2~\mbox{GeV}) = 0.50 ~ (6)$ \cite{Horsley:2015vla} by $\simeq 2.4$ standard deviations.

Using the experimental value for the charged/neutral kaon mass splitting, $\left[ M_{K^+} - M_{K^0} \right]^{exp} = -3.934 ~ (20) ~ \mbox{MeV}$~\cite{PDG}, one gets
 \be
     \left[ M_{K^+} - M_{K^0} \right]^{QCD} (\overline{\mathrm{MS}}, 2~\mbox{GeV}) = -6.00 ~ (15) ~\mbox{MeV} ~ .
     \label{eq:kaon_qcd}
 \ee
 
In order to estimate the light-quark mass difference $(\widehat{m}_d - \widehat{m}_u)$ from the result (\ref{eq:kaon_qcd}) we need to compute the {\it IB slope} (see Eq.~(\ref{eq:MPS_QCD})) $\left[ M_{K^+} - M_{K^0} \right]^{IB}$. Our determination at the physical pion mass and in the continuum and infinite volume limits is
\be
    \left[ M_{K^+} - M_{K^0} \right]^{IB}  =  -2.54 ~ (10)_{stat+fit} ~ (15)_{syst} = -2.54 ~ (18) ~ .
    \label{eq:kaon_qcd_result}
\ee
Putting together the results (\ref{eq:kaon_qcd}) and (\ref{eq:kaon_qcd_result}) with Eq.~(\ref{eq:MPS_QCD}), we get
\be
    \left[\widehat{m}_d - \widehat{m}_u\right](\overline{\mathrm{MS}}, 2~\mbox{GeV}) = 2.380 ~ (87)_{stat+fit} ~ (155)_{syst} ~ (41)_{qQED} ~ \mbox{MeV} = 2.380 ~ (182) ~ \mbox{MeV} ~ ,
   \label{eq:deltamud}
\ee
which is consistent with the previous ETMC determination $2.67 ~ (35)$ MeV \cite{Carrasco:2014cwa} at $N_f = 2+1+1$ and with the recent BMW result, converted in the ($\overline{\mathrm{MS}}, 2~\mbox{GeV}$) scheme, $2.40 ~ (12)$ MeV~\cite{Fodor:2016bgu} at $N_f = 2+1$.
Combining the result (\ref{eq:deltamud}) with our ETMC determination of the average up/down quark mass $m_{ud}(\overline{\mathrm{MS}}, 2~\mbox{GeV}) = 3.70 ~ (17) ~ \mbox{MeV}$ from Ref.~\cite{Carrasco:2014cwa}, we can also compute the $u$- and $d$-quark masses
\be
      \label{eq:mu}
      \widehat{m}_u(\overline{\mathrm{MS}}, 2~\mbox{GeV}) = 2.50 ~ (15)_{stat+fit} ~ (8)_{syst} ~ (2)_{qQED} ~ \mbox{MeV} = 2.50 ~ (17) ~ \mbox{MeV} ~ ,
\ee
\be
      \label{eq:md}
      \widehat{m}_d(\overline{\mathrm{MS}}, 2~\mbox{GeV}) = 4.88 ~ (18)_{stat+fit} ~ (8)_{syst} ~ (2)_{qQED} ~ \mbox{MeV} = 4.88 ~ (20) ~ \mbox{MeV}
\ee
and the ratio
\be
    \left[ \widehat{m}_u / \widehat{m}_d \right]~ (\overline{\mathrm{MS}}, 2~\mbox{GeV}) = 0.513 ~ (18)_{stat+fit} ~ (24)_{syst} ~ (6)_{qQED} = 0.513 ~ (30) ~ ,
    \label{eq:mumd}
\ee
which are consistent within the uncertainties with the current FLAG estimates~\cite{FLAG} at $N_f = 2+1+1$, based on the ETMC results of Ref.~\cite{Carrasco:2014cwa}, and with the recent BMW results~\cite{Fodor:2016bgu} at $N_f = 2+1$. 

Finally, using the ETMC result $m_s(\overline{\mathrm{MS}}, 2~\mbox{GeV}) = 99.6 ~ (4.3) ~ \mbox{MeV}$~\cite{Carrasco:2014cwa} we can obtain a determination of the flavor symmetry breaking parameters $R$ and $Q$, namely
\be
    \label{eq:R}
    R (\overline{\mathrm{MS}}, 2~\mbox{GeV}) \equiv \frac{m_s - m_{ud}}{\widehat{m}_d - \widehat{m}_u} (\overline{\mathrm{MS}}, 2~\mbox{GeV}) = 40.4 ~ (3.3) ~ ,
\ee
\be
    \label{eq:Q}
    Q (\overline{\mathrm{MS}}, 2~\mbox{GeV}) \equiv \sqrt{\frac{ m_s^2 - m_{ud}^2}{ \widehat{m}_d^2 - \widehat{m}_u^2}} (\overline{\mathrm{MS}}, 2~\mbox{GeV}) = 23.8 ~ (1.1) ~ ,
\ee
which are consistent within the errors with the current FLAG estimate $R = 35.6 ~ (5.1)$ and $Q = 22.2 ~ (1.6)$~\cite{FLAG} as well as with the recent BMW results $R = 38.20 ~ (1.95)$ and $Q = 23.40 ~ (64)$~\cite{Fodor:2016bgu}.

\begin{SCfigure}[1.0]
\centering
\includegraphics[scale=0.3]{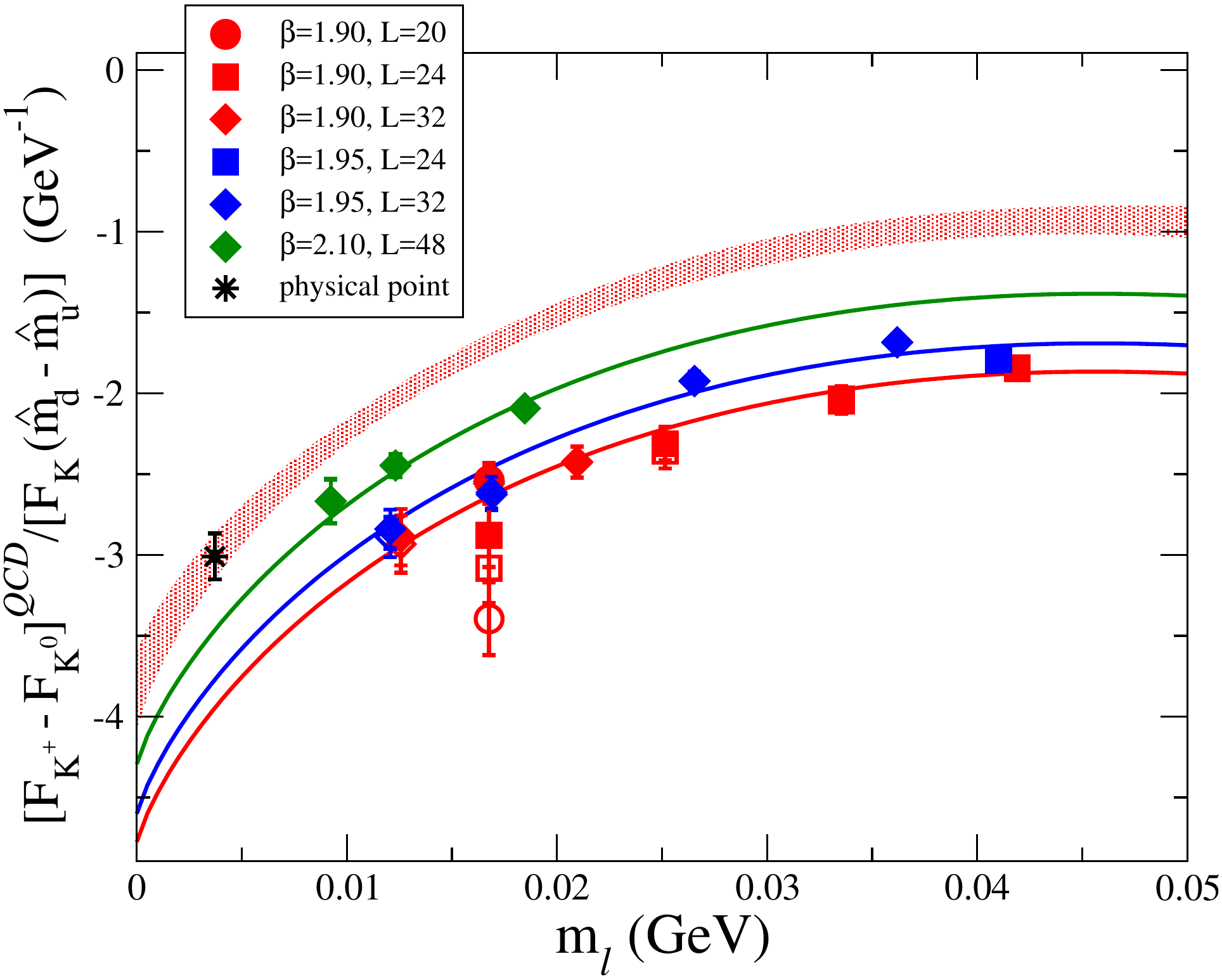}
\caption{\it \footnotesize Results for the IB correction $\left[ F_{K^+} - F_{K^0} \right]^{QCD} / \left[ F_K~(\widehat{m}_d - \widehat{m}_u) \right]$ versus the renormalized light-quark mass $m_\ell$. The empty markers correspond to the lattice data, while the filled ones represent the data corrected for the FVEs obtained in the fitting procedure (\ref{eq:fK_fit}). The solid lines correspond to the results of the combined fit (\ref{eq:fK_fit}) obtained in the infinite volume limit at each value of the lattice spacing. The black asterisk represents the result extrapolated at the physical pion mass $m_\ell = m_{ud} = 3.70 (17)~\mbox{MeV}$ and to the continuum limit, while the red area indicates the corresponding uncertainty as a function of $m_\ell$ at the level of one standard deviation.}
\label{fig:fK_figure}
\end{SCfigure}

It is possible to extract the leading strong IB corrections to the kaon decay constants by studying the ratio of the correlators (see Fig.~(\ref{fig:diagrams}e))
\be
\frac{\delta C^{S_\ell}_K(t)}{C^{(0)}_K(t)} \equiv \frac{\gdsi}{\gdsl} ~ ,
\ee
with the red line representing the strange quark propagator. The IB correction $\delta F_K / F_K$ can be computed by extracting the slope and intercept from the large time behavior of $\delta C^{S_\ell}_K(t) / C^{(0)}_K(t)$ (see Eq.~(\ref{eq:ratio})) according to
\be
\frac{\delta F_K}{F_K} \equiv \frac{1}{\widehat{m}_d - \widehat{m}_u}\frac{\left[ F_{K^+} - F_{K^0} \right]^{QCD}}{F_K} = \frac{1}{m_s + m_\ell} + \frac{\delta Z^{S_\ell}_K}{Z_K} - 2~\frac{\delta M^{S_\ell}_K}{M_K} ~ ,
\ee
where $F_K$ is given by $F_K = \left( m_s + m_{\ell} \right)~Z_K / M^2_K$.
The lattice data have been fitted according to the following ansatz:
\be
\frac{\delta F_K}{F_K} = A + B ~ \frac{\overline{M}^2}{16 \pi^2 f^2_0} + C ~ \frac{\overline{M}^2}{16 \pi^2 f^2_0} \log\left( \frac{\overline{M}^2}{16 \pi^2 f^2_0} \right) + D ~ a^2 + F ~ \frac{\overline{M}^2}{16\pi^2 f_0^2} \frac{e^{-\overline{M} L}}{(\overline{M} L)^{3/2}} ~ ,
\label{eq:fK_fit}
\ee
where $\overline{M}^2 \equiv 2 B_0 m_\ell$, $B_0$, $f_0$ are the QCD low-energy constants at leading order and $A$, $B$, $C$, $D$, $F$ are free parameters to be determined by the fitting procedure. In Eq.~(\ref{eq:fK_fit}) the chiral extrapolation is based on the SU(3) ChPT formulae of Ref.~\cite{Gasser:1984gg} expanded as a power series in terms of the quantity $m_\ell / m_s$, while FVEs are described by a phenomenological term inspired by the leading FVE correction in QCD kaon decay constant in the $p$-regime ($\overline{M} L \gg 1$) \cite{Gasser:1986vb}.
The results of the fitting procedure are shown in Fig.~\ref{fig:fK_figure}.

\noindent At the physical pion mass and in the continuum and infinite volume limits we obtain
\be
\delta F_K / F_K = -3.08 ~ (16)_{stat+fit} ~ (19)_{disc} ~ (10)_{chir} ~ (1)_{FVE} ~ \mbox{GeV}^{-1} = -3.08 ~ (27) ~ \mbox{GeV}^{-1} ~ ,
\label{eq:fK_result}
\ee
where i) $()_{stat+fit}$ indicates the statistical uncertainty including also the one induced by the fitting procedure; ii) $()_{disc}$ is the uncertainty due to discretization effects estimated by including or excluding the term proportional to $a^2$ in Eq.~(\ref{eq:fK_fit}); iii) $()_{chir}$ is the error coming from including the term proportional to the chiral log or substituting it with a quadratic term in $m_\ell$ and iv) $()_{FVE}$ is the uncertainty obtained including or excluding the FVE term in Eq.~(\ref{eq:fK_fit}).

\noindent By using the determination of the light-quark mass difference (see Eq.~(\ref{eq:deltamud})) and the result (\ref{eq:fK_result}) we get the following estimate
\be
\left[ \frac{F_{K^+} - F_{K^0}}{F_K} \right]^{QCD} (\overline{\mathrm{MS}}, 2~\mbox{GeV}) = -0.00730 ~ (48)_{stat+fit} ~ (8)_{disc} ~ (22)_{chir} ~ (6)_{FVE} ~ (12)_{qQED} ~ .
\ee
At order ${\cal O} (\widehat{m}_d - \widehat{m}_u)$, thanks to the fact that pions don't get strong IB corrections, we have
\be
\left[ \frac{F_{K^+} / F_{\pi^+}}{F_K / F_\pi} -1 \right]^{QCD} (\overline{\mathrm{MS}}, 2~\mbox{GeV}) = -0.00365 ~ (28) ~ ,
\ee
a value that is higher (by about 2 standard deviations) than the estimate obtained in Ref.~\cite{Cirigliano:2011tm} by using ChPT, namely
\be
\left[ \frac{F_{K^+} / F_{\pi^+}}{F_K / F_\pi} -1 \right]^{\chi pt} = -0.0022 ~ (6)
\ee
and in agreement with (and more precise than) our previous determination at $N_f = 2$ \cite{deDivitiis:2013xla}.

The violation of the Dashen's theorem for the neutral kaon mass can be represented by the quantity $\epsilon_{K^0}$ defined as~\cite{FLAG}
\be
    \epsilon_{K^0} = \left[\delta M^2_{K^0}\right]^{QED} {\Large \mbox{/}} ~ \left( M_{\pi^+}^2 - M_{\pi^0}^2 \right) ~ .
    \label{eq:epsilon_k0}
\ee
At the physical pion mass and in the continuum limit we obtain
\be
     \epsilon_{K^0} (\overline{\mathrm{MS}}, 2~\mbox{GeV}) = 0.154 ~ (14)_{stat+fit} ~ (20)_{syst} ~ (10)_{qQED} = 0.154 ~ (26) ~ .
    \label{eq:epsilon_K0_result}
\ee
Our result (\ref{eq:epsilon_K0_result}) is in agreement with (and more precise than) both the estimate quoted by FLAG, namely $\epsilon_{K^0} = 0.3 ~ (3)$~\cite{FLAG}, and the recent QCDSF/UKQCD result $\epsilon_{K^0}(\overline{\mathrm{MS}}, 2~\mbox{GeV}) = 0.2 ~ (1)$~\cite{Horsley:2015vla}.

Using the RM123 approach we also address the evaluation of the leading-order e.m.~and strong IB corrections to the $D$-meson mass splitting ($M_{D^+} - M_{D^0}$), and the first lattice determination of the leading-order e.m.~corrections to the $D_s$-meson mass $M_{D_s^+}$.
In the case of $D$-meson mass splitting  we make use of the determination (\ref{eq:deltamud}) of the $u$- and $d$-quark mass difference obtained in the kaon sector to evaluate the strong IB correction and therefore to predict the physical mass splitting ($M_{D^+} - M_{D^0}$) on the lattice.
Within the quenched QED approximation, the QED and QCD contributions to the $D$-meson mass splitting turn out to be
\be
    \label{eq:Dmeson_qed_result}
    \left[ M_{D^+} - M_{D^0} \right]^{QED} (\overline{\mathrm{MS}}, 2~\mbox{GeV}) = 2.42 ~ (22)_{stat+fit} ~ (44)_{syst} ~ (12)_{qQED} ~ \mbox{MeV} = 2.42 ~ (51) ~\mbox{MeV}~ ,
 \ee
 \be
    \label{eq:Dmeson_qcd_result}
    \left[ M_{D^+} - M_{D^0} \right]^{QCD}(\overline{\mathrm{MS}}, 2~\mbox{GeV})  =  3.06 ~ (27)_{stat+fit} ~ (7)_{syst} ~ \mbox{MeV} = 3.06 ~ (27) ~ \mbox{MeV} ~ . 
\ee
Thus, putting together the results (\ref{eq:Dmeson_qed_result}) and (\ref{eq:Dmeson_qcd_result}) we get the prediction
\be
       M_{D^+} - M_{D^0} = 5.47 ~ (30)_{stat+fit} ~ (42)_{syst} ~ (12)_{qQED} ~ \mbox{MeV} = 5.47 ~ (53) ~ \mbox{MeV} ~ ,
   \label{eq:deltaMD}
\ee
which is consistent with the experimental value $M_{D^+} - M_{D^0} = 4.75 (8)$ MeV \cite{PDG} and with the unquenched QED estimate $M_{D^+} - M_{D^0} = 4.68 ~ (16) ~ \mbox{MeV}$ from the BMW collaboration \cite{Borsanyi:2014jba} at $N_f = 1+1+1+1$ within $\simeq 1.4$ standard deviations.

\noindent Finally, our determination of the leading-order e.m.~correction to the $D_s$-meson mass $M_{D_s^+}$ is
\be
    \delta M_{D_s^+} = 5.54 ~ (11)_{stat+fit} ~ (46)_{syst} ~ (28)_{qQED} ~ \mbox{MeV} = 5.54 ~ (55) ~  \mbox{MeV} ~ .
    \label{eq:Dsmeson_result}
\ee

\begin{center}
------------------------------------------------------
\end{center}
We warmly thank R.~Frezzotti and G.C.~Rossi for fruitful discussions. We gratefully acknowledge the CPU time provided by PRACE under the project Pra10-2693 and by CINECA under the initiative INFN-LQCD123 on the BG/Q system Fermi at CINECA (Italy). V.L., G.M., S.S., C.T.~thank MIUR (Italy) for partial support under the contract PRIN 2015. G.M.~also acknowledges partial support from ERC Ideas Advanced Grant n. 267985 ``DaMeSyFla''.

\end{document}